# Alignment-Tolerant Fi-Wi-Fi Free-Space Optical Bridge

**Florian Honz(1), Aina Val Martí(1), Philip Walther(2), Hannes Hübel(1) and Bernhard Schrenk(1)**
(1)AIT Austrian Institute of Technology, Center for Digital Safety & Security / Security & Communication Technologies, 1210 Vienna, Austria.
(2)University of Vienna, Faculty of Physics, 1090 Vienna, Austria
Author e-mail address: florian.honz@ait.ac.at

**Abstract:** We demonstrate a simplified out-door FSO link with modal split for down-/uplink and confirm its long-term stability without active beam tracking. We further prove the duality of modal and directional split through penalty-free full-duplex transmission. 

## 1. Introduction

Free-space optical (FSO) communication is a natural extension of the unprecedented fiber capacity over wireless channels. In contrast to RF-based wireless links that leverage elevated mm-wave and THz frequencies, optical implementations enjoy a ~3 order-of-magnitude higher carrier frequency and low-complexity baseband modulation schemes, which have resulted in fiber-grade transmission capacities of more than 14 Tb/s [1]. FSO links are mainly used in point-to-point configurations, where they serve the permanent or temporary front- and backhaul connectivity of remote radio heads in fiber-scarce radio access network deployments for rural communities [2]. They can further establish a bandwidth-transparent Fi-Wi-Fi bridge over challenging terrain such as rivers and can provide a high-capacity pipe between ground stations and satellites to feed spaceborne communication networks [3, 4]. Despite the offerings of FSO communication in terms of transmission capacity, the wireless transmission channel is heavily impacted by weather and turbulence in the mid-infrared wavelength range [5], where system integration can build on well-established low-cost component technology known to be highly reliable and highly performant. Even though a rate-adaptive transmission layout is thinkable in presence of moderate power fading [6], the highly directional nature of FSO links and the strong susceptibility to angular pointing errors inherent to thermal and mechanical perturbation necessitates active and very accurate beam pointing and tracking to ensure an excellent fiber-to-fiber coupling between the two wireless demarcation points of the FSO channel [7]. This tops the complexity of FSO implementations and hampers their widespread adoption due to unfavorable expenditures for practical deployment.

In this work, we propose a greatly simplified FSO link based on a modal split for down- and uplink transmission, realized through a simple double-clad fiber (DCF) coupler instead of resorting to complex active beam tracking methods. We confirm the improved coupling stability through evaluation of an out-door Fi-Wi-Fi bridge over one month and compare the results to traditional single-mode coupling. We further prove the simultaneous integration of a directional split function through the DCF coupler by demonstrating penalty-free full-duplex signal transmission.

## 2. Alignment-Tolerant FSO Link with Directional Modal Split

A DCF provides multiple guiding regions by introducing a fiber cross-section where two distinct concentric areas are having different refractive indices. We exploit such a DCF configuration for which the single-mode core is used as the launch path in transmit direction, while the inner cladding supports a simultaneous multi-mode propagation modality for signal reception, as shown in

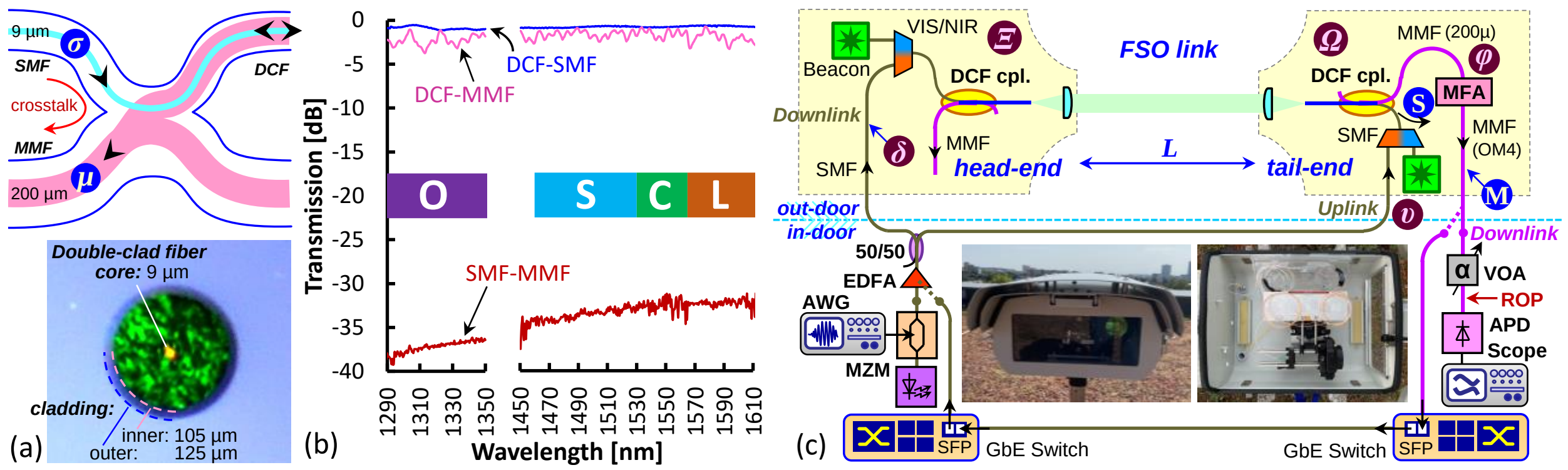

Fig. 1. (a) DCF coupler as directional Fi-Wi interface and (b) its transmission. (c) Experimental setup for the FSO link and installed terminals.

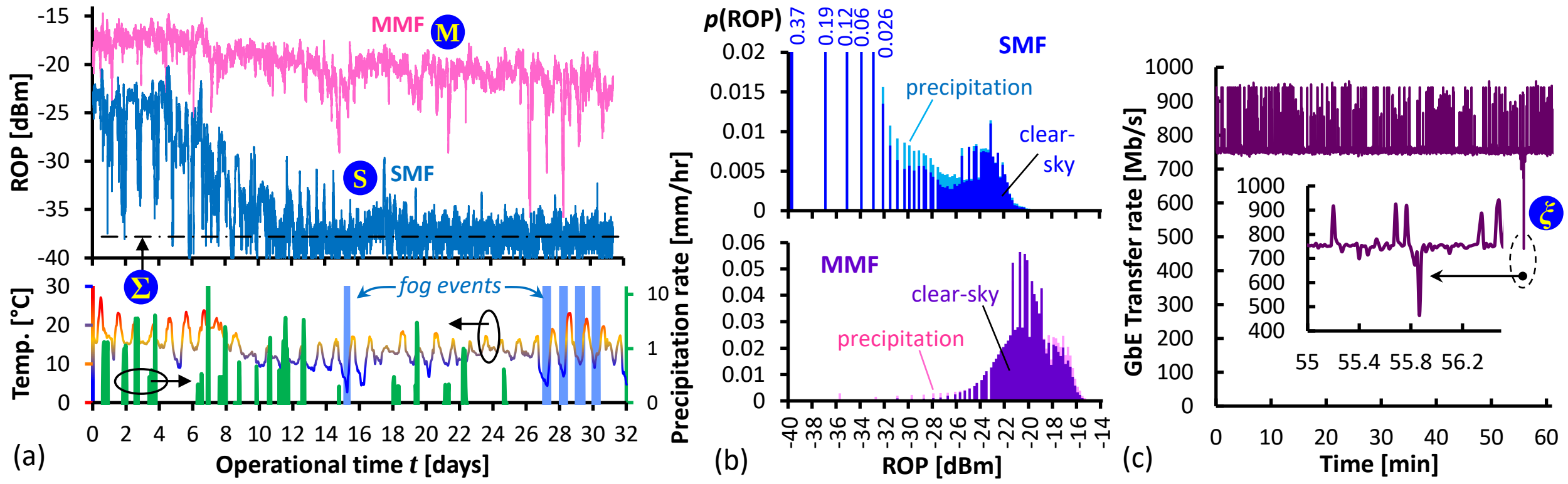

Fig. 2. (a) Long-term measurement of ROP for SMF and MMF coupling and (b) corresponding histograms. (c) GbE transfer rate over FSO link.

Fig. 1a. Implementing a bidirectional layout within a single DCF as fiber-to-wireless demarcation point allows for a well-defined single-mode beam launch and benefits from a large numerical aperture (NA) and surface area for improved light collection at the same time, without compromising the compactness of the air interface. A low-loss directional split between transmit and receive path is accomplished through a DCF coupler, which achieves *(i)* adiabatic single-mode capture from the mode-matched single-mode fiber (SMF, $\sigma$) feeder to the DCF and *(ii)* extraction of the multi-mode signal traversing the inner 105-µm cladding of the DCF into a dedicated multi-mode fiber (MMF, $\mu$). The SMF had a core diameter of 9 µm, a NA of 0.12 and a cut-off wavelength of 1250 nm. It is mode-matched to ITU-T G.652 compatible SMF. The step-index MMF matches the refractive index of the outer DCF cladding. It had a core diameter of 200 µm and a NA of 0.22. Figure 1a includes a cross-section of the DCF, where SMF (core) and MMF (inner cladding) are fed by red and green light, respectively.

Figure 1b reports the transmission of the DCF coupler. The transmission between the DCF and the SMF / MMF ports is -0.9 / -1.5 dB at 1550 nm and is flat across many wavebands. The crosstalk between the (transmit) SMF port and the (receive) MMF port is -33.1 dB at 1550 nm, rendering the DCF coupler a suitable directional split element.

## 3. Experimental Setup for Out-Door FSO Link and Long-Term Stability

The experimental setup of the alignment-tolerant FSO link is presented in Fig. 1c. Two FSO terminals were installed as out-door units ($\Xi$, $\Omega$) at a roof-top location with a spacing of $L$ = 63 m. The DCF coupler constitutes the directional split at the FSO units as described earlier. A mode-field adapter was included at the 200-µm multi-mode port ($\varphi$) to reduce the coupling loss to 50-µm OM4 trunk fiber towards the lab facility. The output beam of the bidirectional DCF at the head-end FSO unit is collimated using a 2-inch lens. Another lens was used for DCF coupling at the tail-end FSO unit. Pointing beacons in the visible-light range assisted the initial face-to-face alignment of the FSO terminals. No further tracking was employed onwards in order to simplify the FSO units.

A transmitter comprised of a Mach-Zehnder modulator (MZM) and a booster EDFA was used to feed down- and uplink ports ($\delta$, $\upsilon$) at both FSO terminals. In this way, the same wavelength and same RF spectrum applies when investigating the impact of a time-decorrelated uplink crosstalk during downlink reception. For the latter, a variable optical attenuator (VOA) precedes an avalanche photodetector (APD) to acquire the data transmission performance as a function of the received optical power (ROP). We used an analogue radio waveform for our evaluation activities since these are touted as especially susceptible to crosstalk noise. The OFDM signal had a RF carrier frequency of $f_C$ = 1.5 GHz, a bandwidth of 250 MHz and consisted of 128 sub-carriers modulated with 64-QAM. Moreover, we employed two GbE switches to test the stability of the FSO link in terms of real-time throughput.

The long-term stability of the FSO link was evaluated over one month at the meteorological transition from summer to autumn. The measurement period included 19 rain and five fog events. The peak-to-peak temperature swing was 24.3°C. Figure 2a presents the ROP at the SMF and MMF outputs of the DCF coupler at the tail-end FSO terminal (S and M in Fig. 1c), together with the temperature and precipitation of a nearby weather station. The SMF coupling shows strong fading effects of >10 dB. After 7 days, it degraded rapidly towards the sensitivity limit of the optical power monitors (Σ). On the contrary, the average in coupled MMF power remained above -21.8 dBm over the entire month. Given the launch power (11 dBm), the mode-field adaptation to OM4 fiber (8.8 dB) and the trunk fiber loss to the lab (5.5 dB), the average FSO loss for the MMF between points $\delta$ and M in Fig. 1c varies from 9.5 ($t$ = 0) to 13.6 dB ($t$ = 30d). Apart from fading effects that are related to inevitable rain and fog events, we attribute the long-term degradation in ROP to sub-optimal fixture of the FSO units for the temporary roof-top installation.

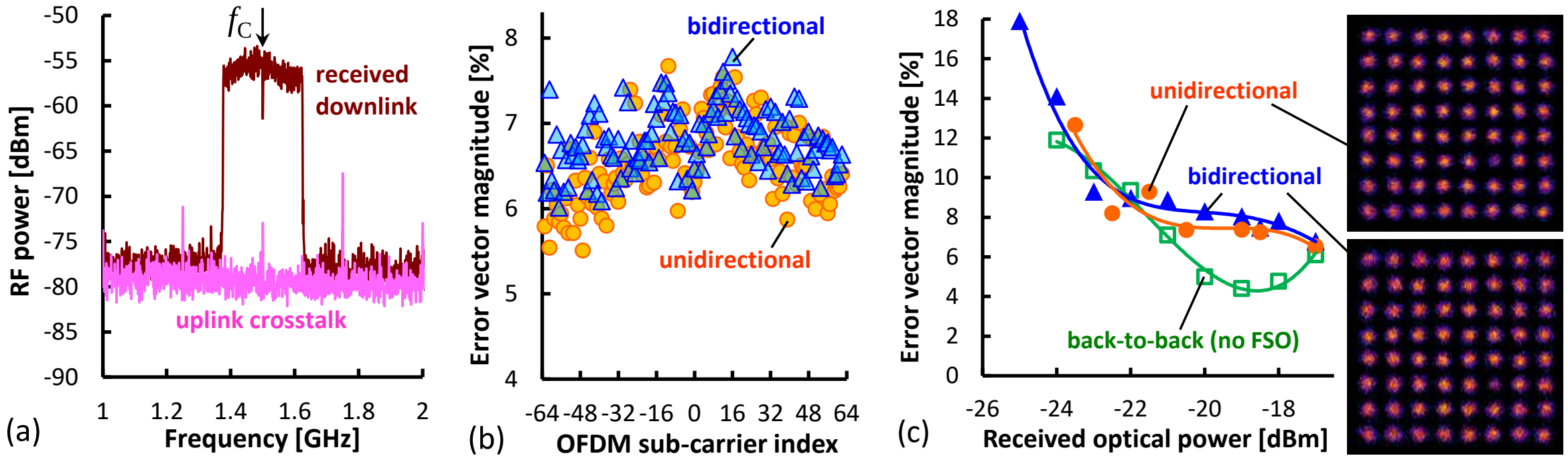

Fig. 3. (a) Received downlink spectrum. (b) Downlink EVM for ROP of -17 dBm. (c) EVM performance as function of ROP.

Figure 2b shows the ROP histograms for SMF and MMF coupling for the entire month and clear-sky conditions. The MMF coupling shows an average ROP of -20.1 dBm with a low spread (3σ = 6.8 dB). ROP values below -28 dBm are due to precipitation and fog. The SMF has a much wider variation in ROP, irrespectively of weather, down to a practical loss of signal. The discretization in ROP towards -40 dBm derives from the employed power monitors.

Fast fluctuations were probed by monitoring the real-time GbE transfer rate between two switches over one hour (Fig. 2c). Besides buffer-related fluctuations between 952 and 744 Mb/s, we noticed a single event ($\xi$) at which the throughput averaged over 1 second dropped to 464 Mb/s. This event is attributed to a wind gust, which we experienced as impacting the fiber coupling due to the exposed temporary roof-top mounting of the FSO terminal.

## 5. Data Transmission Performance over Fi-Wi-Fi Bridge

We then evaluated the reception of an OFDM signal over the FSO link. The received downlink OFDM spectrum at a ROP of -17 dBm is presented in Fig. 3a for the bidirectionally operated FSO link. The crosstalk due to the uplink at the same wavelength and the same OFDM carrier frequency $f_C$, recorded for a disconnected downlink feed, is below the sensitivity of the downlink APD receiver. This proves that the DCF coupler performs the directional split. Optical and RF resources can thus be efficiently re-used for down- and uplink data transmission.

Figure 3b reports the downlink EVM as function of the OFDM sub-carrier for a ROP of -17 dBm. Without uplink, meaning that the FSO link is operated unidirectionally for the sole purpose of downlink transmission (○), the average EVM is 6.5%. When the uplink is activated and the FSO link becomes bidirectional (△), the EVM increases marginally by 0.3% at this ROP, which corresponds to an optical signal-to-crosstalk ratio of 22 dB.

To investigate the EVM penalty due to a reduced signal-to-crosstalk ratio at low ROP, Figure 3c presents the EVM over ROP for the unidirectional (●) and bidirectional (▲) FSO link. For the latter, the EVM increases for ROP levels below -21 dBm and reaches 10% at a ROP of -22.8 dBm (▲). However, as the EVM for the unidirectional link without uplink reveals, this increase leads to a similar ROP of -22.7 dBm at 10% (●) and is therefore attributed to the sensitivity limit of the APD receiver rather than to uplink crosstalk noise. Comparison is further made to the back-to-back case without FSO link (□). The lower EVM of 4.4% at a ROP around -19 dBm is explained by the absence of mode partition noise arising due to the coupling to an expanded 200-µm MMF and mode-field adaptation to OM4 fiber. We noticed saturation effects of the APD towards a ROP of -17 dBm, leading to an EVM degradation.

## 6. Conclusion

We have proposed a simplified FSO link exploiting a modal split for down- and uplink directions. By employing a DCF as fiber-to-wireless demarcation point, efficient and alignment-robust coupling can be accomplished, as shown for a temporary out-door installation: Evaluation of the FSO link over one month rendered the optical coupling as robust, despite the purely passive nature of the FSO terminals without active beam tracking. Apart from a higher coupling efficiency, the peak-to-peak spread in ROP has been reduced by 10.6 dB. Full-duplex data transmission showed no penalty despite re-using the same wavelength and OFDM carrier frequency for radio signal relay over the Fi-Wi-Fi bridge in down- and uplink direction, rendering the proposed modal split as an additional directional split.

Acknowledgement: This work was supported by the ERC under the EU Horizon-2020 programme (grant agreement No 804769).